\documentclass{article}

\usepackage[final]{neurips_2023_ml4ps}

\usepackage[utf8]{inputenc} 
\usepackage[T1]{fontenc}    
\usepackage{hyperref}       
\usepackage{url}            
\usepackage{booktabs}       
\usepackage{amsfonts}       
\usepackage{nicefrac}       
\usepackage{microtype}      
\usepackage{xcolor}         

\usepackage{graphicx}
\usepackage{subfigure}

\title{HIDM: Emulating Large Scale HI Maps using Score-based Diffusion Models}

\author{%
  Sultan Hassan~\thanks{NASA Hubble Fellow}\\
  New York University\\
  Flatiron Institute \\
  University of the Western Cape\\
  \texttt{sultan.hassan@nyu.edu} \\
  \And
  Sambatra Andrianomena \\
  South African Radio Astronomy Observatory\\
  Department of Physics \& Astronomy \\ University of the Western Cape\\
  \texttt{andrianomena@gmail.com} \\
}

\begin{document}

\maketitle

\begin{abstract}
 Efficiently analyzing maps from upcoming large-scale surveys requires gaining direct access to a high-dimensional likelihood and generating large-scale fields with high fidelity, which both represent major challenges. Using CAMELS simulations, we employ the state-of-the-art score-based diffusion models to simultaneously achieve both tasks. We show that our model, HIDM, is able to efficiently generate high fidelity large scale HI maps that are in a good agreement with the CAMELS's power spectrum, probability distribution, and likelihood up to second moments. HIDM represents a step forward towards maximizing the scientific return of future large scale surveys.
\end{abstract}

\section{Introduction}
Imaging the neutral Hydrogen (HI) distribution -- a promising tracer of dark matter and large scale structure -- over cosmological scales is a primary target of several ongoing and upcoming surveys such as the Square Kilometer Array~\citep[SKA,][]{2013ExA....36..235M}, the Hydrogen Epoch of Reionization Array~\citep[HERA,][]{2017PASP..129d5001D}, the Low Frequency Array~\citep[LOFAR,][]{2013A&A...556A...2V} the Vera C. Rubin Observatory Legacy Survey of Space and Time~\citep[LSST,][]{2019ApJ...873..111I}, Nancy Grace Roman Space Telescope~\citep[Roman,][]{2015arXiv150303757S}, Spectro-Photometer for the History of the Universe, Epoch of Reionization, and Ices Explorer~\citep[SPHEREx,][]{2014arXiv1412.4872D}, and  Euclid~\citep[][]{2016SPIE.9904E..0OR}.  A key challenge to analyzing outputs from these large scale surveys relies in the high-dimensionality, which poses several significant challenges including memory requirement and computational expense. For instance, traditional large scale simulations are computationally too expensive to enable field level inference via standard techniques such as Markov chain Monte Carlo (MCMC) samplers. To effectively analyze any large scale high-dimensional observed maps (including HI), access to the likelihood and ability to generate new synthetic examples conditioned on parameters are both required.

To achieve these goals, previous studies have used various techniques of generative models, such as variants of Generative Adversarial Networks~\citep[e.g. ][]{2019arXiv190412846Z,2021ApJ...915...71V,2021ApJ...916...42W,2023arXiv230307473A}, Variational Autoencoders~\citep[e.g.][]{2021MNRAS.504.5543L,2021MNRAS.503.4446C},  and flow-based models~\citep[e.g.][]{2022ApJ...937...83H,2022arXiv221112724F,2022MNRAS.516.2363D}. However, most of these techniques fail to fully capture the high frequencies (i.e. small scale features), where non-linearity is higher, especially when applied on high resolution large scale maps. To maximize the scientific return of future surveys, an efficient and flexible deep learning model is needed to correctly model features over all scales. Currently, diffusion models achieve the state-of-the-art performance in image generation tasks\footnote{https://paperswithcode.com/method/diffusion}. In addition, some previous works~\citep{2022arXiv221103812A,2022arXiv221112444M} have shown that diffusion models are successful in generating posterior samples of realistic astrophysical fields.  The general flow of these models is based on the idea of gradually injecting noise to convert the signal to a white noise, and then generating new examples by gradually removing back the noise~\citep[e.g.][]{2020arXiv201113456S}.  In addition and similar to the flow-based models, obtaining the likelihood is possible via a simple change of variables. Hence, developing diffusion models to generate large scale HI maps would be very useful to effectively analyze future surveys.

\section{Training dataset}
\begin{figure*}
\subfigure{\includegraphics[scale=0.6]{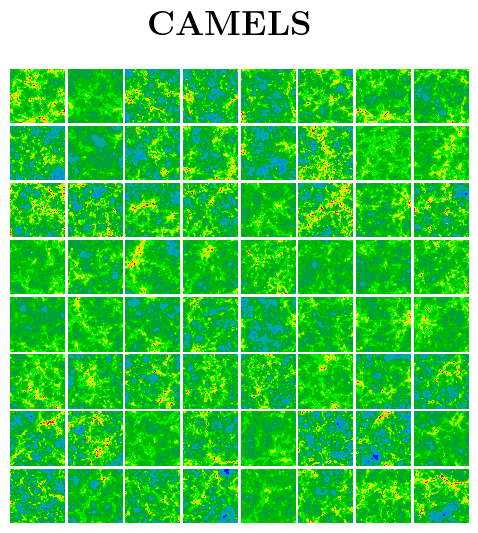}}
\hfill 
\subfigure{\includegraphics[scale=0.6]{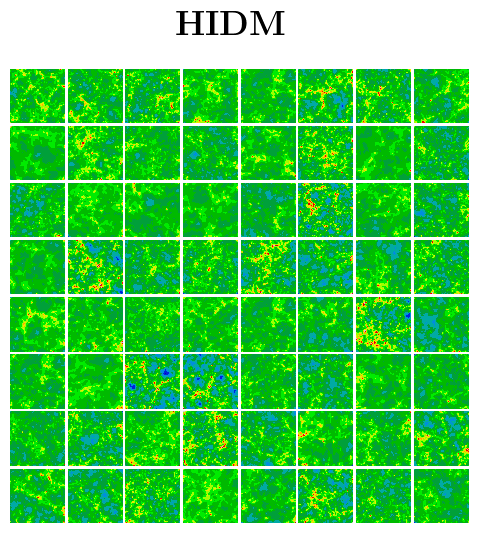}}
\caption{Comparison between HIDM and CAMELS in terms of 64 random generated examples of large scale HI maps. Both simulations visually show  indistinguishable diversity of the HI large scale fluctuations.  These maps cover an area of 25 $h^{-1}$ cMpc $\times$ 25 $h^{-1}$cMpc with 64 pixels on a side, resulting in a resolution of $\sim$ 0.4 $h^{-1}$ cMpc. The color scale in these maps show the column density range $\log_{10}$ N$_{\rm HI}$/cm$^{-2}$=14-22. }
\label{fig:maps}
\end{figure*}

We use the publicly available HI maps from CAMELS Mulitfield Dataset~\citep[CMD,][]{CMD} an outcome of the CAMELS project~\citep{2021ApJ...915...71V}. Through the variation of several cosmological and astrophysical parameters, the resulting maps from CAMELS simulations show a large diversity of HI fluctuations on cosmological scales as shown in the left panel of Figure~\ref{fig:maps}. The HI maps considered in this study is from the IllustrisTNG simulations~\citep{2017MNRAS.465.3291W, 2018MNRAS.473.4077P} at $z=6$ with 64$\times$64 pixels and size of 25 ${\rm Mpc}/h$, where $h=0.6711$.

\begin{figure*}

\subfigure{\includegraphics[scale=0.4]{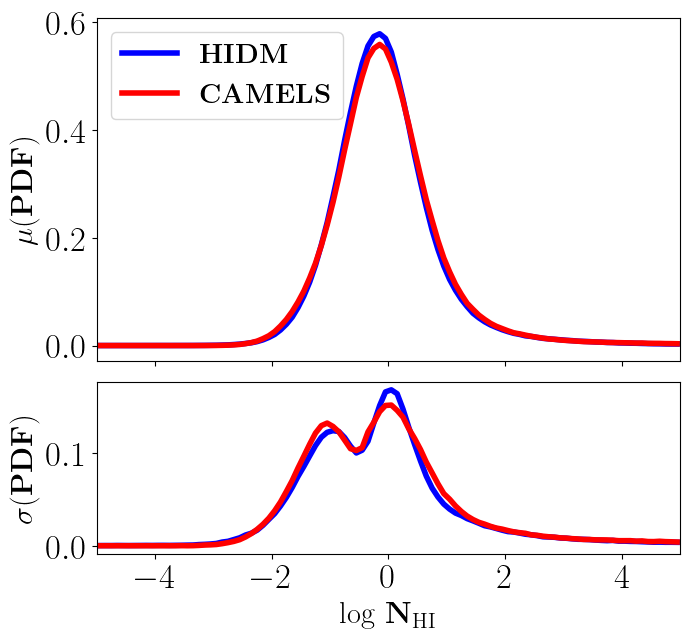}}
\hfill 
\subfigure{\includegraphics[scale=0.4]{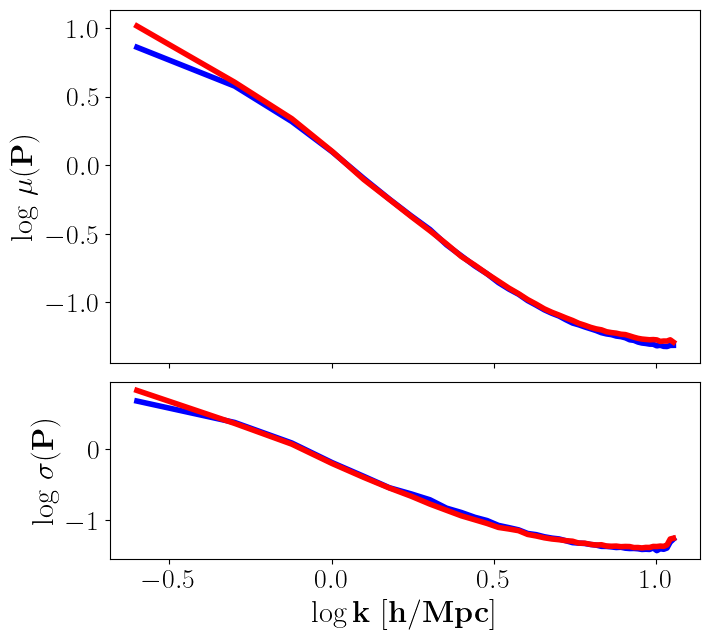}}
\caption{Comparison between HIDM (blue) and CAMELS (red) in terms of the mean ($\mu$) and standard deviation ($\sigma$) of the power spectrum (left) and probability distribution of HI large scale maps. Both simulations show similar statistics up to second moments.}
\label{fig:stats}
\end{figure*}

\section{Score-based diffusion models}
We follow closely the implementation by~\citet{2020arXiv201113456S} since it provides a somewhat faster sampling through solving a Stochastic Differential Equation (SDE) as well as access to the high dimensional likelihood. We here briefly describe the basic framework of diffusion models. As mentioned earlier, the diffusion process gradually corrupts data into random noise over time via the following SDE:
\begin{equation}
  \rm   dx = f(x,t) dt + g(t) dw,\,
\end{equation}
where $\rm f(x,t)$ is the drift coefficient, $\rm g(t)$ is diffusion coefficient, and $\rm dw$ is a standard Brownian motion. In this process, the distribution of data $P_{0}$ is gradually converted into a tractable prior distribution $P_{T}$ over time. Due to the large nature of the diffusion processes,  $P_{T}$ would have no memory of the initial $P_{0}$, and hence this provides essential tests for out-of-distribution detection~\citep[e.g.][]{2022arXiv221107740G}.  The score function -- the gradient of the prior $\nabla_{x} \log P_{T}$ -- arises in the reverse process to generate samples via the following reverse-time SDE:
\begin{equation}
\rm dx  = [f(x,t) - g^{2}(t)  \nabla_{x} \log P_{t}] dt + g(t) dw.\,
\end{equation}
The score $\nabla_{x} \log P_{t}$ is estimated using a time-dependent score model (S) which is approximated by a standard U-Net architecture to minimize the following loss function $\mathcal{L}$:
\begin{equation}
    \mathcal{L} =E_{x(0)}E_{x(t)} [|| S(x(t),t) -   \nabla_{x(t)} \log P_{t} (x(t)| x(0)) ||^{2}_{2}]\,.
\end{equation}

In our case, we gradually perturb HI maps to the prior distribution assuming the following SDE: $\rm dx=\lambda^{t} dw$, where $t\in[0,1]$. This corresponds to a standard deviation ($\sigma$) of the marginal distribution $P_{t}(x(t)|x(0))$ of $\frac{1}{2\log\lambda}(\lambda^{2t}-1)$ that gradually converts HI maps to white noise.  The resulting loss function in this case reads:
\begin{equation}
    \mathcal{L} = \frac{1}{N}\sum_{N} ( S(x+\sigma z,t) + z)^{2},\,
\end{equation}
where $z$ are random samples from a normal distribution i.e. $\mathcal{N}(0,1)$. 

For reproducibility, our U-Net architecture  consists of four convolutional blocks for down and up sampling with $3\times 3$ 64 filters in each layer that is conditioned on time (t). We use Adam optimizer with learning rate of 10$^{-3}$ and a diffusion coefficient $\lambda=25$ for a total of 500 training epochs (convergence achieved from first 100 epochs) with a mini batch size of 128. To generate new samples, one needs to solve the reverse-time SDE, which in our case reads:
\begin{equation}
\rm dx = -\frac{1}{2}\lambda^{-2t} S(x,t) dt.\,    
\end{equation}
In particular, we start from a random white noise $P(T)$ and integrate this equation in the reverse direction from $t = T\, {\rm to}\, 0$ to generate new realization of HI maps $P(0)$. It is then straightforward to compute the exact likelihood with change of variable formula (similar to flow-based models) as follows:
\begin{equation}
  \rm  \log P(0)= \log P(T) - \frac{1}{2}  \int_{0}^{T} \frac{d \lambda^{2}}{dt} {\rm div}\, S(x(t),t)\, dt,\,
\end{equation}
where div S is the divergence of the score function representing the trace of the Jacobian of the change of variable.
\section{Results}
Using the process explained in the previous section, we design HI map diffusion model (HIDM). Figure~\ref{fig:maps} shows 64 random samples from CAMELS (left) versus HIDM (right). This figure demonstrates the ability of HIDM to generate high quality synthetic examples of large scale HI maps with diversity that is indistinguishable from the target (CAMELS). 
To quantify the image generation quality by HIDM, we consider the probability distribution function (PDF, left) and power spectrum (P, right) as shown in Figure~\ref{fig:stats}. We first select 1,000 random realizations from CAMELS and similarly generate 1,000 random examples from HIDM. We then compute the PDF over all the 1,000 realizations from each model, and measure the mean PDF ($\mu(\rm PDF)$) and standard deviation PDF ($\sigma(\rm PDF)$). We repeat the same procedure to compute the mean power ($\mu(\rm P)$) and standard deviation power ($\sigma(\rm P)$) out of the 1,000 random realizations. We show the corresponding statistics of CAMELS in red and HIDM in blue. It is evident that HIDM fully captures the PDF and power spectra up to the second moment (variance) over all 1,000 random realizations, albeit a minor difference in the large scale power. This shows the ability of HIDM to efficiently emulate the behaviour of CAMELS simulations. While a more comprehensive comparison in terms of higher order statistics might be needed to validate HIDM against CAMELS, it is already challenging for surveys to measure the average power spectrum due to the large observational systematics~\citep[e.g.][]{2022ApJ...925..221A}. It is worth noting that sampling these 1,000 realizations from HIDM takes 20 seconds on a single GPU as compared to running 1,000 CAMELS simulations with each simulation run can take up to approximately $\sim$ 48 hours on 24 CPU node. This is at least a factor of 10$^{6}$ speed gain as compared to CAMELS simulations. This speed again will enable performing inference at the field level in future surveys. 

We additionally compute the likelihood over all 1,000 realizations and find $\rm \log P(\rm CAMELS) = 21.5\pm 0.7$ bits/dim, whereas the $\rm \log P(\rm HIDM) = 20.9 \pm 0.4$ bits/dim. These likelihood values are consistent within 1-$\sigma$ level. This shows that HIDM has successfully learned the likelihood of HI maps from the CAMELS simulations, and hence this model can efficiently perform implicit inference (i.e. simulation based inference) as soon as conditioning on cosmological and astrophysical parameters is implemented, which we leave to future works. 


\section{Conclusion}
We have introduced HIDM, a generative model of high fidelity HI maps using score-based diffusion models trained on samples from the state-of-the-art CAMELS simulations. We have demonstrated the model ability to generate high quality synthetic examples of large scale HI maps with diversity that is indistinguishable from the target (see Figure~\ref{fig:maps}). We have also shown, over 1,000 random samples, the model fully captures the CAMELS' PDF and power spectra, up to the second moment (variance, see Figure~\ref{fig:stats}). By applying a simple change of variables, the model enables an exact likelihood computation. We have then shown that the model has successfully learned the target likelihood within 1-$\sigma$ level. 

In this initial attempt, we have verified the ability of score-based diffusion process to successfully model large scale HI maps and provide access to high dimensional likelihood. To maximize the use of this model, conditioning on the various cosmological and astrophysical parameters is required in order to enable field level inference. Conditioning on parameters can be easily implemented as an additional vector embedding next to the time information within the time-dependent U-Net architecture (the Score model). We leave developing the conditional score model on parameters to future works. The HI maps used in this study do not account for various instrumental effects such as foregrounds, thermal noise and angular resolution, which might reduce the applicability of HIDM to analyze maps from future surveys. We leave to future works adding all instrumental effects to perform accurate forecasting to future HI surveys. In addition, we plan to test the scalability of diffusion models as a function of resolution and dimension. The efficiency of HIDM is quite remarkable as compared to CAMELS simulations, achieving at least a factor of 10$^{6}$ speed gain, which will enable performing accurate inference at the field level, and hence maximizing the scientific return of future large scale surveys. 

\section{Broader impact}
HIDM represents a step forward towards efficiently analyzing big data expected from large scale surveys. By accurately emulating large scale fields from simulations, HIDM could enable field-level inference, once conditioned on parameters, to minimize loss of complex information due to data compression down to summary statistics. The same analysis can be applied to large scale maps of different observables such as Carbon or Lyman-$\alpha$ emissions, which are the target for many of the upcoming intensity mapping surveys. These surveys will carry a wealth of information that will eventually constrain our understanding of galaxy evolution, formation and large scale structure, and hence the impact of developing emulators of large scale fields, such as HIDM, is wide enough to influence different communities in cosmology and astronomy.

\begin{ack}
SH acknowledges support for Program number HST-HF2-51507 provided by NASA through a grant from the Space Telescope Science Institute, which is operated by the Association of Universities for Research in Astronomy, incorporated, under NASA contract NAS5-26555. SH also acknowledges support through NYU Postdoctoral Research and Professional Development Support Grants and Simons Foundation. SA acknowledges financial support from the South African Radio Astronomy Observatory (SARAO).  
\end{ack}

\bibliographystyle{unsrtnat}
\bibliography{ref}


\end{document}